\documentclass[conference]{IEEEtran}
\usepackage{booktabs}
\usepackage{amsmath,amssymb,amsfonts}
\usepackage{algorithm}
\usepackage{algpseudocode}
\usepackage{graphicx}
\usepackage{xcolor}
\begin{document}
%
\title{Distributed Learning for Time-varying Networks: \\ A Scalable Design}
\author{\IEEEauthorblockN{Jian ~Wang\IEEEauthorrefmark{1}, Yourui ~Huangfu\IEEEauthorrefmark{1}, Rong ~Li\IEEEauthorrefmark{1}, Yiqun ~Ge\IEEEauthorrefmark{2}, Jun ~Wang\IEEEauthorrefmark{1}}
\IEEEauthorblockA{\IEEEauthorrefmark{1}Wireless Technology Laboratory, Huawei Technologies, Hangzhou, China}
\IEEEauthorblockA{\IEEEauthorrefmark{2}Ottawa Research Center, Huawei Technologies, Ottawa, Canada}
Emails: \{wangjian23, huangfuyourui, lirongone.li, yiqun.ge, justin.wangjun\}@huawei.com
}

\maketitle
\IEEEpeerreviewmaketitle
\begin{abstract}
The wireless network is undergoing a trend from ``connection of things'' to ``connection of intelligence''. With data spread over the communication networks and computing capability enhanced on the devices, distributed learning becomes a hot topic in both industrial and academic communities. Many frameworks, such as federated learning and federated distillation, have been proposed. However, few of them takes good care of obstacles such as the time-varying topology resulted by the characteristics of wireless networks. In this paper, we propose a distributed learning framework based on a scalable deep neural network (DNN) design. By exploiting the permutation equivalence and invariance properties of the learning tasks, the DNNs with different scales for different clients can be built up based on two basic parameter sub-matrices. Further, model aggregation can also be conducted based on these two sub-matrices to improve the learning convergence and performance. Finally, simulation results verify the benefits of the proposed framework by compared with some baselines.
\end{abstract}

\IEEEpeerreviewmaketitle

\section{Introduction}
\label{sec1}
Future wireless systems are envisaged to be deeply integrated with artificial intelligence (AI) technologies to provide both communication and AI services \cite{wang2020artificial}. In the past decade, the great progress in three key driving forces, i.e., diverse data sets, advanced algorithms and powerful computing capability, has played an important role in the success of AI applications \cite{russell2016artificial}. Hence, the deployment of AI in a particular scenario such as wireless systems always begins with the considerations with three factors, i.e., how to collect data, how to process data, and by which device to process. For example, in scenarios of Internet of Things (IoT) and vehicular communication systems, a large volume of data will be collected and analyzed by devices to provide classification, regression, prediction, and decision making functions to handle different tasks. In these cases, distributed learning is a proper enabler for AI solutions \cite{chen2021distributed}.

A lot of distributed learning frameworks have been recently proposed, within which the most popular one is federated learning (FL) \cite{mcmahan2017communication}. In FL, each client trains a local learning model using its local collected data, and then uploads the local model to a central server. The central server aggregates multiple local models to a global one and broadcasts it to the clients for next round training. FL exploits the computing capability of distributed clients and keeps users' data local at clients without direct exchanging. A commonly used assumption for FL is that the basic structure of the learning models (both local ones and the global one) are identical. Although model compression, sparsification and pruning may alleviate the computation and communication overhead, they must be based on a same structure.

On the other hand, distributed clients may need to cooperate with each other to finish a task together. The learning made by each client is not only depended on its local data, but also related to the data and actions of its neighbors. Informations are allowed to exchanged in this case, and the input of the learning model on each client may include both local and neighboring data. The numbers of neighbors for different clients may be different and change over time, which implies that the input of the learning model for each client may be different and time-varying. Retraining a model repetitively is impractical due to the large overhead during the training phase. Pre-defining a maximum possible number of neighbors and training a corresponding large model may be a solution. However, the assumption of knowing the maximum possible number of neighbors is not always feasible in practice, and this model can not handle the situation when the number of neighbors exceeds the pre-defined one.

In this paper, we propose a novel distributed learning framework which can be adopted in the wireless networks with time-varying topology. First of all, to cope with different input dimensions of the learning models at clients, a scalable deep neural network (DNN) structure is designed. Here, similar to previous works \cite{guo2020structure,wang2021smart}, the permutation equivalence (PE) and permutation invariance (PI) properties, which can be found in many applications \cite{zaheer2017deep}, are considered. Based on this PE/PI prior knowledge of applications, the DNN structure can be simplified and viewed as a combination of two basic blocks. Then, the DNNs with different scales can be easily built up through using different numbers of basic blocks at clients to handle both local and neighboring data. Moreover, the model aggregation at the central server can be taken out also based on the two basic blocks, which makes the model aggregation on DNNs with different scales possible. We evaluate the proposed framework through simulations and show that with less model parameters, the proposed framework outperforms the design with a fixed size.

The paper is organized as follows. The application scenario considered in this paper are introduced in Section \ref{sec2}. Then, in Section \ref{sec3}, the proposed distributed learning framework is elaborated. Simulation results and discussions are provided in Section \ref{sec4}. Finally, we summarize the work in Section \ref{sec5}.

\section{System Model}
\label{sec2}
\begin{figure}[t]
	\centering
	\includegraphics[width=\columnwidth]{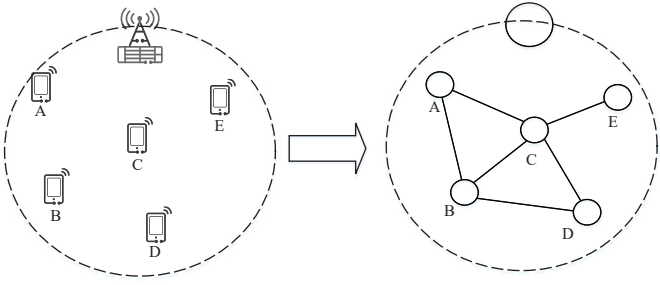}
	\caption{System model.}
	\label{fig:systemmodel}
\end{figure}
As shown in Fig. \ref{fig:systemmodel}, we consider a wireless network with one central unit such as a base station (BS) and multiple distributed clients such as mobile user equipments (UEs). A distributed learning task is assigned to this wireless network, where each of the distributed clients need to perform the learning based on both its local data and the data from its neighbors. A typical application scenario of this system model is object recognition where different cameras and sensors capture different features from the same object, which can be regarded as a multi-view learning task \cite{xu2013survey}. This system model can also find its application in IoT systems and wireless sensor networks (WSNs).

We characterize the system by a graph, where the vertexes are the distributed clients and the edges show the connectivity between clients. Different clients may have different numbers of neighbors, e.g., Client E has only one neighbor of Client C, while Client D has two neighbors of Client B and D. Moreover, as a wireless network, the topology of the graph is of high dynamics. Due to the quality of wireless link and on/off activities of clients, the graph structure may be time-varying. For example, considering the graph in Fig. \ref{fig:systemmodel}, client C has four neighbors of Client A, B, D and E in first time slot. If, in the second time slot, Client D departs from the system, the neighbors of Client C will change to Client A, B and E. The DNNs deployed on all the clients need to handle both local and neighboring data, so at least the input dimensions of these DNNs are different. As the graph changing over time, this input dimension of the DNN on the same client may also be time varying.

A server is deployed on the central unit of the system. It can help with speeding up the training progress through aggregating local models from the clients. However, the model aggregation must take into consideration the situation when the scales of the DNNs are different.

\section{Distributed Learning with a Scalable DNN Design}
\label{sec3}
\subsection{PE/PI-based Scalable DNN Design}
\label{sec3.1}
PE and PI are well-known and widely-used properties which can be elaborated as follows. Given an input vector $\mathbf{X}=\left[\mathbf{x}_{1}, \mathbf{x}_{2}, \ldots, \mathbf{x}_{k}\right]$, if for an arbitrary permutation $\pi(k)=p_{k}$ on $\mathbf{X}$, i.e., $\tilde{\mathbf{X}}=\left[\mathbf{x}_{p_{1}}, \mathbf{x}_{p_{2}}, \ldots, \mathbf{x}_{p_{K}}\right]$, the output of function $\mathbf{Y}=f(\mathbf{X})=\left[\mathbf{y}_{1}, \mathbf{y}_{2}, \ldots, \mathbf{y}_{k}\right]$ follows the same permutation, i.e., $\tilde{\mathbf{Y}}=f(\tilde{\mathbf{X}})=\left[\mathbf{y}_{p_{1}}, \mathbf{y}_{p_{2}}, \ldots, \mathbf{y}_{p_{K}}\right]$, we say that the function $f$ is PE for the input $\mathbf{X}$. And if the output of function $f$ does not change even with the permutation of the input $\mathbf{X}$, we say that it has the PI property. Many distributed learning tasks have PE and PI properties. For example, in the UE scheduling problem in wireless networks, the scheduling priority of the target UE depends on both its own state such as channel condition and the states of other UEs. However, permuting the states of UEs before inputting them into the scheduler, the output scheduling priorities of the UEs permute in the same way. The PE and PI properties can also be found in multi-view learning task, where the permutation in the order of input neighboring data may not affect the output of the learning model.

\begin{figure}[t]
	\centering
	\includegraphics[width=\columnwidth]{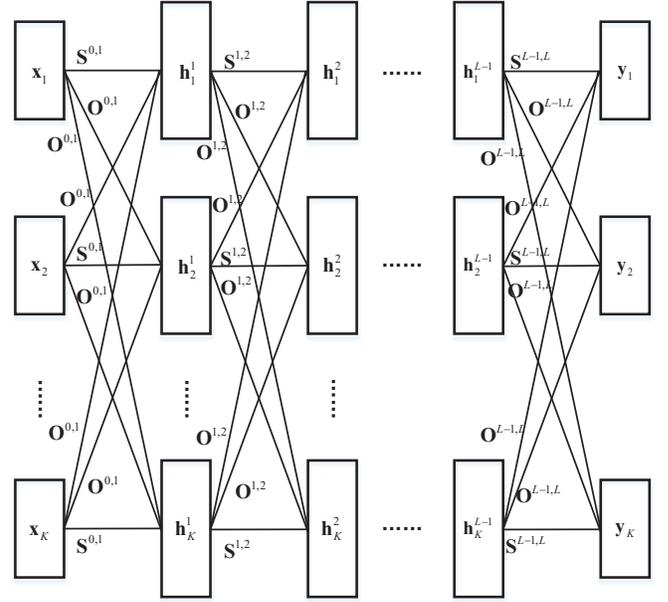}
	\caption{Structure of DNNs with PE/PI property.}
	\label{fig:NN}
\end{figure}

To realize the aforementioned function $f$ through a DNN with $L$ layers, we can rely on the parameter sharing structure as shown in Fig. \ref{fig:NN}. The input $\left[\mathbf{x}_{1}, \mathbf{x}_{2}, \ldots, \mathbf{x}_{K}\right]$ includes states from $K$ clients, i.e., one target client and its $K-1$ neighbors. ${\bf{h}}_k^l, k=1,2,\ldots,K, l=1,2,\ldots,L$ is the output vector on the $l$-th hidden layer. If $\left[\mathbf{y}_{1}, \mathbf{y}_{2}, \ldots, \mathbf{y}_{k}\right]$ is considered as the output, the DNN is of PE property. The permutation on the input $\left[\mathbf{x}_{1}, \mathbf{x}_{2}, \ldots, \mathbf{x}_{K}\right]$ results in the same permutation on the output $\left[\mathbf{y}_{1}, \mathbf{y}_{2}, \ldots, \mathbf{y}_{K}\right]$. If only $\mathbf{y}_{\rm{target}}, {\rm{target}} \in \{1, 2, \ldots, K\}$ is considered, the DNN is of PI property because the permutation on the input will not change the output. The processing of the $l$-th hidden layer can be expressed as
\begin{equation}\label{eq:eq1}
\tilde{\mathbf{h}}^{l}=\sigma\left(\tilde{\mathbf{W}}^{l-1, l} \tilde{\mathbf{h}}^{l-1}+\mathbf{b}^{l}\right),
\end{equation}
where $\tilde{\mathbf{h}}^{l} = [{\bf{h}}_1^l,{\bf{h}}_1^l,\ldots, {\bf{h}}_K^l ]$, $\tilde{\mathbf{W}}^{l-1, l}$ is the weight matrix and $\mathbf{b}^{l}$ is the bias vector of the $l$-th hidden layer, and $\sigma(\cdot)$ is the activation function. With ${{\bf{W}}^{l - 1,l}} = {\rm{ }}\left[ {\begin{array}{*{20}{c}}
{{{\widetilde {\bf{W}}}^{l - 1,l}}}&{{{\bf{b}}^l}}
\end{array}} \right]$ and ${{\bf{h}}^l} = {\left[ {\begin{array}{*{20}{c}}
{{{\widetilde {\bf{h}}}^l}}&{\bf{1}}
\end{array}} \right]^{\rm{T}}}$, where $\bf{1}$ is a vector with all-one elements and length same to ${\bf{h}}_k^l$, Eq. \ref{eq:eq1} can be rewritten as
\begin{equation}\label{eq:eq2}
\mathbf{h}^{l}=\sigma\left(\mathbf{W}^{l-1, l} \mathbf{h}^{l-1}\right).
\end{equation}
To guarantee the PE/PI property of the DNN, $\mathbf{W}^{l-1, l}$ should have the following structure
\begin{equation}\label{eq:eq3}
{{\bf{W}}^{l - 1,l}} = \left[ {\begin{array}{*{20}{c}}
{\begin{array}{*{20}{c}}
{{{\bf{S}}^{l - 1,l}}}&{{{\bf{O}}^{l - 1,l}}}\\
{{{\bf{O}}^{l - 1,l}}}&{{{\bf{S}}^{l - 1,l}}}
\end{array}}&{\begin{array}{*{20}{c}}
 \ldots &{{{\bf{O}}^{l - 1,l}}}\\
 \ldots &{{{\bf{O}}^{l - 1,l}}}
\end{array}}\\
{\begin{array}{*{20}{c}}
 \vdots & \vdots \\
{{{\bf{O}}^{l - 1,l}}}&{{{\bf{O}}^{l - 1,l}}}
\end{array}}&{\begin{array}{*{20}{c}}
 \ddots & \vdots \\
 \cdots &{{{\bf{S}}^{l - 1,l}}}
\end{array}}
\end{array}} \right],
\end{equation}
where $\mathbf{W}^{l-1, l}$ consists of two sub-matrices, i.e., $\mathbf{S}^{l-1, l}$ and $\mathbf{O}^{l-1, l}$, which is also shown in Fig. \ref{fig:NN}. Hence, DNNs with different scales can be built up based on these two sub-matrices to handle different input dimensions.

\subsection{The Proposed Distributed Learning Framework}

\begin{figure}[t]
	\centering
	\includegraphics[width=\columnwidth]{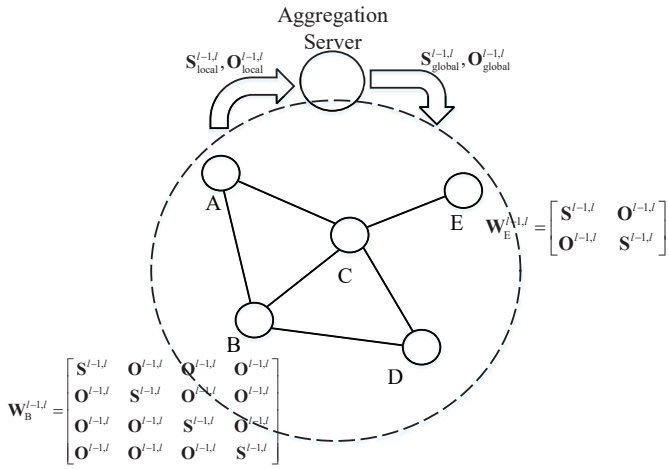}
	\caption{Proposed distributed learning framework.}
	\label{fig:DLframework}
\end{figure}

Relying on the PE/PI-based scalable DNN design, we propose a distributed learning framework as shown in Fig. \ref{fig:DLframework}. Each distributed client uses a DNN with the aforementioned structure, whose scale can be adjusted according to the number of neighbors. For example, Client B has three neighbors, hence the parameter matrix of its DNN can be expressed as
\begin{equation}\label{eq:eq4}
{\bf{W}}_{\rm{B}}^{l - 1,l} = \left[ {\begin{array}{*{20}{c}}
{\begin{array}{*{20}{c}}
{{{\bf{S}}^{l - 1,l}}}&{{{\bf{O}}^{l - 1,l}}}\\
{{{\bf{O}}^{l - 1,l}}}&{{{\bf{S}}^{l - 1,l}}}
\end{array}}&{\begin{array}{*{20}{c}}
{{{\bf{O}}^{l - 1,l}}}&{{{\bf{O}}^{l - 1,l}}}\\
{{{\bf{O}}^{l - 1,l}}}&{{{\bf{O}}^{l - 1,l}}}
\end{array}}\\
{\begin{array}{*{20}{c}}
{{{\bf{O}}^{l - 1,l}}}&{{{\bf{O}}^{l - 1,l}}}\\
{{{\bf{O}}^{l - 1,l}}}&{{{\bf{O}}^{l - 1,l}}}
\end{array}}&{\begin{array}{*{20}{c}}
{{{\bf{S}}^{l - 1,l}}}&{{{\bf{O}}^{l - 1,l}}}\\
{{{\bf{O}}^{l - 1,l}}}&{{{\bf{S}}^{l - 1,l}}}
\end{array}}
\end{array}} \right].
\end{equation}
However, Client E has only one neighbor and its parameter matrix is
\begin{equation}\label{eq:eq5}
{\bf{W}}_{\rm{E}}^{l - 1,l} = \left[ {\begin{array}{*{20}{c}}
{{{\bf{S}}^{l - 1,l}}}&{{{\bf{O}}^{l - 1,l}}}\\
{{{\bf{O}}^{l - 1,l}}}&{{{\bf{S}}^{l - 1,l}}}
\end{array}} \right].
\end{equation}

When the topology of the wireless network changes, the scale of the DNN used at each client can also change accordingly. For example, if Client B departs from the network at the $t$-th time slot, the parameter matrix of the DNN used by Client A will change from ${\bf{W}}_{{\rm{A}},\left( {t - 1} \right)}^{l - 1,l}$ to ${\bf{W}}_{{\rm{A}},\left( {t} \right)}^{l - 1,l}$, where
\begin{equation}\label{eq:eq6}
{\bf{W}}_{{\rm{A}},\left( {t - 1} \right)}^{l - 1,l} = \left[ {\begin{array}{*{20}{c}}
{{\bf{S}}_{\left( {t - 1} \right)}^{l - 1,l}}&{{\bf{O}}_{\left( {t - 1} \right)}^{l - 1,l}}&{{\bf{O}}_{\left( {t - 1} \right)}^{l - 1,l}}\\
{{\bf{O}}_{\left( {t - 1} \right)}^{l - 1,l}}&{{\bf{S}}_{\left( {t - 1} \right)}^{l - 1,l}}&{{\bf{O}}_{\left( {t - 1} \right)}^{l - 1,l}}\\
{{\bf{O}}_{\left( {t - 1} \right)}^{l - 1,l}}&{{\bf{O}}_{\left( {t - 1} \right)}^{l - 1,l}}&{{\bf{S}}_{\left( {t - 1} \right)}^{l - 1,l}}
\end{array}} \right],
\end{equation}
and
\begin{equation}\label{eq:eq7}
{\bf{W}}_{{\rm{A}},\left( {t} \right)}^{l - 1,l} = \left[ {\begin{array}{*{20}{c}}
{{\bf{S}}_{\left( {t} \right)}^{l - 1,l}}&{{\bf{O}}_{\left( {t} \right)}^{l - 1,l}}\\
{{\bf{O}}_{\left( {t} \right)}^{l - 1,l}}&{{\bf{S}}_{\left( {t} \right)}^{l - 1,l}}
\end{array}} \right].
\end{equation}

After training the local DNN based on the local and neighboring data, each client can upload the local training results ${\bf{S}}_{{\rm{local}}}^{l - 1,l}$ and ${\bf{O}}_{{\rm{local}}}^{l - 1,l}$ to the central server, where global sub-matrices, i.e., ${\bf{S}}_{{\rm{global}}}^{l - 1,l}$ and ${\bf{O}}_{{\rm{global}}}^{l - 1,l}$, are obtained through model aggregation. All commonly used aggregation algorithms can be adopted here. Although the scales of the DNNs used by all the clients may be different from each other, through the same basic building blocks, i.e., the two sub-matrices, model aggregation can be done. Similar as in FL, the model aggreation can make used of data from multiple clients and speed up the whole learning procedure.

$\mathbf{S}^{l-1, l}$ and $\mathbf{O}^{l-1, l}$ are of the same size, which is supposed to be $a \times b$, then the total number of parameters of the DNN is $aK \times bK = abK^2$, where $K$ is the number of neighbors plus one. Thanks to the parameter sharing, the number of trainable parameters is only $2ab$, which is also the communication payload size for local model uploading and global model broadcasting. Compared to directly using a large DNN with $abK^2$ parameters, the DNN used in the proposed distributed learning framework can save up to $K^2/2$ times computation and communication overhead.

\section{Simulations}
\label{sec4}
To evaluate the performance of the proposed distributed learning framework, we design a test case and show the simulation results in this section.

\begin{figure}[t]
	\centering
	\includegraphics[width=\columnwidth]{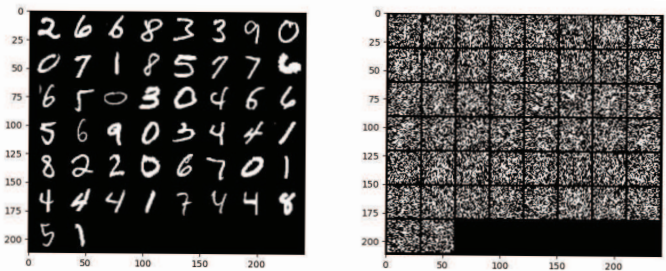}%
	\caption{Sample pictures in MNIST (left:original, right: -10dB noisy).}
	\label{fig:mnist}
\end{figure}

We consider a classification task on the MNIST \cite{lecun2010mnist} in a network with 5 distributed clients. Both the training set with 60000 samples and the test set with 10000 samples are divided into five non-overlapping subsets, each of which is focused by one of the five clients. To turn the task into a cooperative one, we add noise to the samples, and different noisy versions of the same picture are allocated to the target client and its neighbors. The target client tries to identify the class of the item in the picture by input its local noisy version and neighboring noisy versions of the picture. The original and noisy version of some sample pictures are shown in Fig. \ref{fig:mnist}.

\begin{figure*}[t!]
	\centering
	\includegraphics[width=1.6\columnwidth]{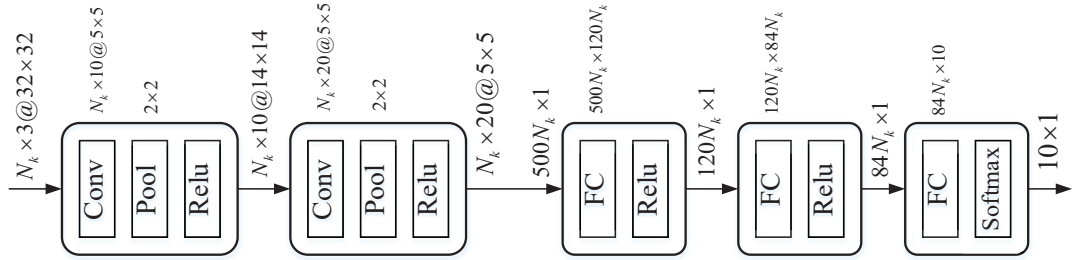}%
	\caption{General structure of the used DNN.}
	\label{fig:structure}
\end{figure*}

For comparison, we consider three baselines. The first one is a one-client scenario with no neighbors and no model aggregation. For the second baseline, cooperation among clients is neglected and each client tries to do the classification based only on its local data, i.e., one noisy figure. For the third baseline, a large DNN is used, where five figures can be processed simultaneously. Obviously, this large DNN is very suitable for Client C who has four neighbors. However, for other clients with fewer neighbors, zeros will be padded to make the input have the same as five pictures.

The proposed framework and the baselines use a similar DNN structure, which includes two convolution layers with 2-by-2 pooling and Relu activation, and three fully connected (FC) layers with Relu activation for the first two FC layers and Softmax activation for the last FC layer. The general structure of the DNN is shown in Fig. \ref{fig:structure}, where $N_k$ is a scaling factor which has different values for different method. For the first and second baselines without considering data from neighbors, we set $N_k=1$. For the third baseline with a large DNN, $N_k$ is set to be 5. Because the DNN used by the proposed method is built up based on two sub-matrices, i.e., $\mathbf{S}^{l-1, l}$ and $\mathbf{O}^{l-1, l}$, both of which have the same dimension as the one used in the first baseline, $N_k$ is equal to the number of neighbors of Client $k$ plus one. It is worth noting that thanks to the parameter sharing design, the number of trainable parameters of the DNN used by the proposed method is only twice the number for the first baseline. Other simulation parameters are summarized in Table \ref{tab:simulationsettings}.

\begin{table}[h]
	\centering
	\caption{Simulation Settings}
	\begin{tabular}{c|c}
		\hline
		Parameters		                &   Values \\
		\hline
		Number of clients	            &	5 \\
        \hline
        Data set                        &   CIFAR10 \\
        \hline
        Number of training samples      &   12000 per client \\
        \hline
        Number of test samples          &   2000 per client  \\
        \hline
        Learning rate                   &   0.05 \\
        \hline
        Decaying rate of learning rate  &   0.99 \\
        \hline
        Training batch size             &   500 \\
        \hline
        Test batch size                 &   128 \\
        \hline
        Number of training epoches      &   20 \\
        \hline
        Testing frequency               &   every 2 epoches \\
		\hline
	\end{tabular}
	\label{tab:simulationsettings}
\end{table}

\begin{figure}[t]
	\centering
	\includegraphics[width=\columnwidth]{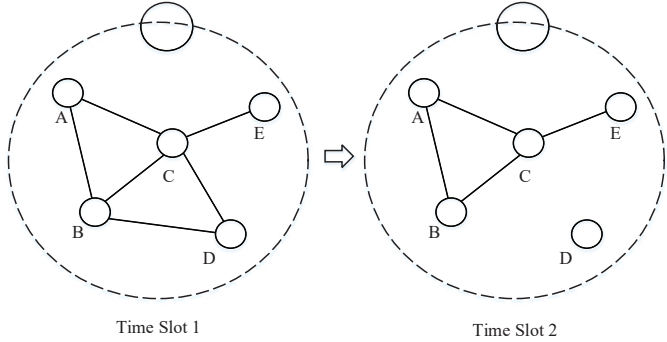}%
	\caption{The simulated time-varying scenario.}
	\label{fig:simscenario}
\end{figure}
We consider the scenario with a time-varying topology as shown in Fig. \ref{fig:simscenario}. In the first time slot, the connectivity among clients is the same as Fig. \ref{fig:systemmodel}. Then, in the second time slot, Client D is off, hence, the numbers of neighbors of Client B and C are decreased by one. The connectivities can be represented as
\begin{equation}\label{eq:eq8}
{I_{{\rm{TS1}}}} = \left[ {\begin{array}{*{20}{c}}
1&1&1&0&0\\
1&1&1&1&0\\
1&1&1&1&1\\
0&1&1&1&0\\
0&0&1&0&1
\end{array}} \right]
\end{equation}
and
\begin{equation}\label{eq:eq9}
{I_{{\rm{TS2}}}} = \left[ {\begin{array}{*{20}{c}}
1&1&1&0&0\\
1&1&1&0&0\\
1&1&1&0&1\\
0&0&0&0&0\\
0&0&1&0&1
\end{array}} \right]
\end{equation}
for the first and seconde time slots, respectively.

\begin{figure}[t]
	\centering
	\includegraphics[width=\columnwidth]{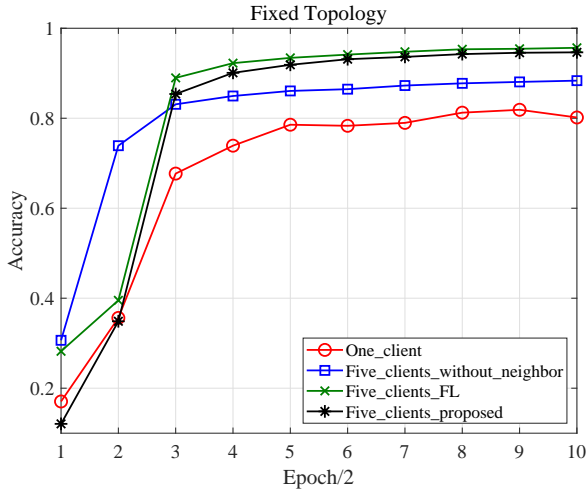}%
	\caption{Accuracy of different methods in fixed topology scenario.}
	\label{fig:fixed}
\end{figure}
The performance comparison among aforementioned methods is firstly done in the fixed topology for the first time slot. The accuracy performances during the training phases of different methods are shown in Fig. \ref{fig:fixed}. The first baseline with only one client considered performs the worst, while with the help of model aggregation, the second baseline performs better than the first one although no data from neighbors is used. With consideration of the neighboring data, both the third baseline and the proposed method show superior performance. Although the proposed method has a little performance degradation compared to the third baseline, the third baseline uses a much larger DNN. In the simulated case, the number of parameters of DNNs used in the third baseline is 1348630, while the number of parameters of DNNs used in the proposed method is only 109628. This means only one tenth the communication and computation overhead of the third baseline is needed for the proposed method. Moreover, the maximum possible number of participants of the learning may be hard to pre-define in practice, which also hindering the third baseline from being deployed in real systems.

\begin{table}[!htbp]
	\centering
	\caption{Inference Performance for Time-varying Scenario}
	\begin{tabular}{c|c|c|c}
		\hline
		\multicolumn{2}{c|}{Method}                &  Baseline 3 (large DNN)  & Proposed method \\
        \hline
        \multicolumn{2}{c|}{Number of Parameters}  &  1348630                 & 109628          \\
		\hline
		    	      &	Client A                    &  0.9589                 & 0.9487          \\ \cline{2-4}
                      &	Client B                    &  0.9668                 & 0.9593          \\ \cline{2-4}
        Time Slot 1   &	Client C                    &  0.9709                 & 0.9656          \\ \cline{2-4}
                      &	Client D                    &  0.9592                 & 0.9499          \\ \cline{2-4}
                      &	Client E                    &  0.9366                 & 0.9208          \\ \cline{2-4}
                      &	Average                     &  0.9585                 & 0.9489          \\
        \hline
		    	      &	Client A                    &  0.9596                 & 0.9490          \\ \cline{2-4}
                      &	Client B                    &  0.9577                 & 0.9493          \\ \cline{2-4}
        Time Slot 2   &	Client C                    &  0.9677                 & 0.9601          \\ \cline{2-4}
                      &	Client E                    &  0.9357                 & 0.9206          \\ \cline{2-4}
                      &	Average                     &  0.9551                 & 0.9447          \\
		\hline
	\end{tabular}
	\label{tab:timevarying}
\end{table}

We now evaluate the performance of the time-varying scenario. Because the first and second baselines do not consider neighboring data, the time-varying topology will not affect their performance. Hence, we only compare the performance between the proposed framework and the third baseline. The DNNs are trained for 20 epoches, and are fixed for inference for 10 iterations. The averaging results over the 10 iterations are shown in Table \ref{tab:timevarying}. It is shown that both methods can work in the time-varying networks. The departure of Client D results in reductions in numbers of neighbors for Client B and C, so the performance of Client B and C degrade. Again, we emphasize that the third baseline is used here for comparison, it is hard to be deployed in practice due to the assumption of knowing the maximum possible participants of the learning.

\section{Conclusions}
\label{sec5}
In this paper, we proposed a distributed learning framework based on a scalable DNN design to conduct cooperative tasks in wireless networks with time-varying topology. Relying on the PE/PI property, parameter matrices of DNNs with different scales are built up based on the same two basic blocks. Hence, distributed clients can form its own DNN based on these two basic sub-matrices according to the number of its neighbors. The changing on the number of neighbors is handled by rebuilding the DNN accordingly. Aggregation of local models with different scales can be done at the central server also based on these two basic sub-matrices, which improves the convergence and performance of the global model. Simulations show that, the proposed method can effectively handle the scenario with time-varying topology. 


{\footnotesize
\bibliographystyle{IEEEtran}
\bibliography{DistPI}
}

\end{document}